
\magnification=\magstep1
\baselineskip=15pt
\overfullrule=0pt

\def\C{{\bf C}}
\def\F{{\cal F}}
\def\G{{\cal G}}
\def\H{{\cal H}}

\def\N{{\cal N}}
\def\O{{\cal O}}
\def\R{{\cal R}}
\def\Z{{\bf Z}}
\def\half{{1 \over 2}}
\def\12{{\scriptstyle{1\over 2}}}
\def\31{{\scriptstyle{1\over 3}}}
\def\23{{\scriptstyle{2\over 3}}}
\def\43{{\scriptstyle{4\over 3}}}

\def\Im{{\rm Im}}
\def\tr{{\rm tr}}
\def\notin{\in \!\!\!\!\! / ~}
\def\next{\hfil\break}

\rightline{UCLA/98/TEP/11}
\rightline{Columbia/Math/98}
\rightline{NSF-ITP-98-062}

\bigskip

\centerline{{\bf SPECTRAL CURVES FOR SUPER-YANG-MILLS WITH ADJOINT}}

\medskip

\centerline{{\bf HYPERMULTIPLET FOR GENERAL LIE ALGEBRAS}
\footnote*{Research supported in part by
the National Science Foundation under grants PHY-95-31023, 
PHY-94-07194 and 
DMS-95-05399.}}

\bigskip
\bigskip
\bigskip

\centerline{{\bf Eric D'Hoker}${}^1$ 
            {\bf and D.H. Phong} ${}^2$}

\bigskip

\centerline{${}^1$ Department of Physics}
\centerline{University of California, Los Angeles, CA 90024, USA;}
\centerline{Institute for Theoretical Physics}
\centerline{University of California, Santa Barbara, CA 93106, USA}

\bigskip

\centerline{${}^2$ Department of Mathematics}
\centerline{Columbia University, New York, NY 10027, USA}

\bigskip
\bigskip
\bigskip

\centerline{\bf ABSTRACT}

\bigskip

The Seiberg-Witten curves and differentials for $\N=2$ supersymmetric 
Yang-Mills theories with one hypermultiplet of mass $m$ in the adjoint 
representation of the gauge algebra $\G$, are constructed for arbitrary
classical  or exceptional $\G$ (except $G_2$). The curves are obtained from 
the recently established Lax pairs with spectral parameter for the (twisted) 
elliptic Calogero-Moser integrable systems associated with the algebra $\G$.   
Curves and differentials are shown to have the proper group theoretic and 
complex analytic structure, and to behave as expected when $m$ tends either 
to 0 or to $\infty$. By way of example, the prepotential for $\G = D_n$, 
evaluated with these techniques, is shown to agree with standard perturbative 
results. A renormalization group type equation relating the prepotential to 
the Calogero-Moser Hamiltonian is obtained for arbitrary $\G$, generalizing a 
previous result for $\G = SU(N)$. Duality properties and decoupling to 
theories with other representations are briefly discussed.

\vfill\break

\centerline{\bf I. INTRODUCTION}

\bigskip

Substantial evidence has accumulated in support of the connection between 
the structure of the low energy effective action of $\N=2$ super-Yang-Mills 
theory in four dimensions [1] and certain integrable systems [2,3,4,5,6,7]. 
For 
reviews, see [8]. Compelling arguments were given on 
general grounds that this connection should hold true [4]. In particular, 
the 
Hitchin system [9] was proposed for $\N=2$ super-Yang-Mills theory with 
gauge algebra $SU(N)$ and one massive hypermultiplet in the adjoint 
representation of $SU(N)$, and the Seiberg-Witten curve 
and differential [4] were naturally obtained from it. A possible relation 
between the spectral curves arising from the Hitchin system and those 
associated with the elliptic Calogero-Moser systems [10,11,12] was 
suggested in [5] and established for the $SU(N)$ gauge algebra by Krichever 
in 
unpublished work. 

\medskip

In a recent paper [7], we showed that the 
Calogero-Moser integrable system indeed captures the physics of the low 
energy dynamics of $\N=2$ supersymmetric Yang-Mills theory with gauge 
algebra $SU(N)$ and with one 
hypermultiplet of mass $m$ in the adjoint representation of $SU(N)$.  We 
checked the perturbative contributions and evaluated 1- and 2-instanton 
corrections with the help of a renormalization group type equation which 
was also established in [7] (see also [13]). Decoupling the full 
hypermultiplet by letting $m\to \infty$ while keeping the vacuum expectation 
values of the gauge scalar 
fixed reproduced the gauge theory without hypermultiplet [14]. By letting 
the mass $m$, as 
well as some of the vacuum expectation values of the gauge scalar tend to 
$\infty$ while tuning their differences, we recovered the gauge theory with
massive  hypermultiplets in the fundamental representation of $SU(N)$ [15]. 
In
special cases, it was possible to get product unitary gauge groups with
hypermultiplets in fundamental and bi-fundamental representations, of the 
type 
solved by Witten using branes, string theory and M-theory [16].

\medskip

In the present paper, we propose Seiberg-Witten curves and associated 
differentials in terms of elliptic Calogero-Moser systems  for $\N=2$
supersymmetric Yang-Mills theories with arbitrary  gauge algebra $\G$
(except $G_2$), and 
with one hypermultiplet of mass $m$ in the adjoint  representation of the 
gauge
algebra. The precise correspondence is with the ordinary elliptic 
Calogero-Moser system when $\G$ is simply laced, and with the {\it twisted  
Calogero-Moser system} when $\G$ is non-simply laced. The latter was 
introduced 
in a companion paper [17], and both will be reviewed below. The modulus 
$\tau$ 
of the elliptic curve $\Sigma$ that underlies the Calogero-Moser systems is
given in terms of the super-Yang-Mills coupling $g$ and theta angle $\theta$ 
by
$$
\tau = { \theta \over 2 \pi} + {4\pi i \over g^2}.
\eqno (1.1)
$$
In view of the ultra-violet finiteness of these theories, this coupling is 
well-defined. In terms of the Lax operators $L(z)$ and $M(z)$ for the 
(twisted) 
elliptic Calogero-Moser system, {\it with spectral parameter} $z \in 
\Sigma$, 
the curve and the differential take the form
$$
\eqalign{
\Gamma \ : \ R(k,z)  & = \det (k I - L(z))=0,\cr
d \lambda & = k dz. \cr}
\eqno (1.2)
$$

\medskip

Until now, the construction of the Seiberg-Witten curve and differential 
from 
the elliptic Calogero-Moser system could be carried out only for $\G=SU(N)$, 
(as in [7]) since it is only for $\G=SU(N)$ that the relevant Lax pair with 
spectral parameter was known [11]. For $\G \not=SU(N)$, the 
situation was as follows. For classical Lie algebras $\G$,
Lax operators without spectral parameter for the elliptic Calogero-Moser 
systems were discovered long ago [12]. However, from the very 
proposal of the spectral curves in (1.2), it is clear that Lax pairs with 
spectral parameter are needed. Though a Lax pair with a free parameter was
introduced by Inozemtsev [18], its dependence on  this parameter appears 
unsuited for Seiberg-Witten theory. For exceptional Lie algebras, no Lax 
pairs (with or without spectral parameter)  appear to be known at all. (See
e.g. [28] for a recent discussion.)

\medskip

In a first companion paper [17], we give an explicit and systematic 
construction of the Lax pairs with spectral parameter for the ordinary (and 
twisted) elliptic Calogero-Moser  systems, associated with any simple Lie 
algebra $\G$, including exceptional ones. In a second companion paper [19], 
we 
show that under certain scaling behaviors of $\tau$ and $m\to \infty$, the 
ordinary (resp. twisted) Calogero-Moser systems tend towards Toda systems 
associated with the untwised affine Lie algebras $\G ^{(1)}$ (resp. dual 
affine 
Lie algebras $(\G ^{(1)})^\vee)$. Using the results of [17] and of [19], we 
shall argue in the present paper that the  spectral curves constructed from 
the 
elliptic Calogero-Moser systems indeed generate the Seiberg-Witten curves 
for 
super-Yang-Mills theory
with one massive adjoint hypermultiplet. The crucial requirements on the 
curve
and differential that we shall check here are as follows. 

\medskip

\item{(a)} The curve $R(k,z)=0$ must be invariant under the action of the 
Weyl
group of $\G$ and have appropriate analytic behavior;
\item{(b)} The differential $d\lambda$ must be meromorphic on the curve 
$R(k,z)=0$, with poles independent of moduli, and residues linear in the 
hypermultiplet mass $m$.
\item{(c)} In the $m=0$ limit, the curve and differential must reduce to 
those
of  the $\N=4$ supersymmetric gauge theory which receives no quantum 
corrections;
\item{(d)} In the $m\to \infty$ limit, while keeping vacuum expectation 
values 
of the gauge scalars fixed, and running the gauge coupling according to the 
renormalization group, the curve and differential for the super-Yang-Mills 
theory without hypermultiplet must be recovered.

\medskip

The remainder of this paper is organized as follows. In \S II, we present 
the
Hamiltonians and Lax pairs of the ordinary and twisted Calogero-Moser 
systems,
obtained in [17]. In \S III, we justify the Seiberg-Witten curves and 
differential, by checking points (a), (b), (c), (d) and (e), discussed 
above, 
using the results of [19].
In \S IV, we obtain by way of example, the spectral curve for $\G=D_n$ in 
the 
weak coupling limit, and show that the effective prepotential obtained in 
this 
limit agrees with the standard result from quantum field theory.
In \S V, we present a number of applications. We generalize to the case of 
an
arbitrary Lie algebra
$\G$ the renormalization group type equation established previously for 
$SU(N)$
in [7],  in which the variation of the prepotential with the gauge coupling  
in
terms of  the Calogero-Moser Hamiltonian. We discuss the duality properties 
for
various $\G$. We indicate how curves for $\N=2$ super-Yang-Mills theories 
with
hypermultiplets in certain other representations of the gauge algebra $\G$ 
may
be obtained from the adjoint hypermultiplet case by suitable decoupling 
limits.

\medskip

Finally, we point out that Seiberg-Witten curves have also been derived by 
using string theory methods. One method is by exploiting the appearence of 
enhanced gauge symmetries at certain singular compactifications. (See for 
example [20].)
A second method is by obtaining supersymmetric Yang-Mills theory as an 
effective theory on a configuration of branes in string theory or M-theory.  
This approach was pioneered in [16], (see also [21]) for $SU(N)$ gauge group 
(and products thereof). The extension to other classical groups is discussed in 
[22].  Relations between the string theory and M-theory approaches and 
integrable systems were proposed 
in
[16,23].

\bigskip
\bigskip

\centerline{\bf II. CALOGERO-MOSER HAMILTONIANS AND LAX OPERATORS}

\bigskip

The elliptic Calogero-Moser integrable systems associated with a 
finite-dimensional simple Lie algebra $\G$ of rank $n$, involve $n$ complex 
degrees of freedom $x_i$ and their canonical momenta $p_i$, $i=1,\cdots ,n$. 
The Hamiltonian is defined by 
$$
H= \half \, p\cdot p - \sum _{\alpha \in \R (\G)} \half m_{|\alpha|} ^2
\wp _{\nu (\alpha)} (\alpha \cdot x).
\eqno (2.1)
$$
Here, $\R (\G)$ is the set of all (non-zero) roots of $\G$, and $m_{|\alpha 
|}$ are complex constants dependent only on the Weyl orbit $|\alpha |$. The 
$\wp _\nu$ are Weierstrass functions defined by
$$
\wp _\nu (u) = \sum _{l=0} ^{\nu -1} \wp (u + 2 \omega _a {l \over \nu}),
\eqno (2.2)
$$
where $\omega _a$ is one of the half periods $\omega _1, \omega _2 $ or 
$\omega _3 = \omega _1 + \omega _2$; for definiteness, we choose $\omega 
_1$.
Finally, the function $\nu (\alpha)$ enters as follows.

\item{(1)}
The {\it ordinary Calogero-Moser system} is defined by $\nu (\alpha ) =1$ 
for 
{\it all roots } of $\G$.
\item{(2)}
The {\it twisted Calogero-Moser system} is defined for non-simply laced $\G$ 
by
\itemitem{} $\nu (\alpha)=1$ for all {\it long roots} of $\G$;
\itemitem{} $\nu (\alpha)=2$ for all {\it short roots} of $B_n,\, C_n$ and 
$F_4$;
\itemitem{} $\nu (\alpha) =3$ for all {\it short roots} of $G_2$.
\item{(3)}
For simply laced $\G$, the twisted and ordinary Calogero-Moser systems 
coincide. 

\medskip

The elliptic Calogero-Moser Hamiltonians are completely integrable in the 
sense 
that there exists a Lax pair of $N\times N$ dimensional matrix valued 
functions 
of $x$ and $p$, denoted by $L$ and $M$, such that the Hamilton-Jacobi 
equations of the elliptic Calogero-Moser system on $x$ and $p$ are 
equivalent 
to the Lax equation, given by
$$
\dot L = [L,M].
\eqno (2.3)
$$
Integrability follows from the existence of the Lax equation, since the 
latter 
automatically guarantees that the quantities $\tr L^{\gamma +1}$ are 
conserved 
integrals of motion for $\gamma =0,\cdots ,\infty$.
In fact, in [17] a stronger result was shown to hold. There exists a pair of 
Lax operators $L(z)$ and $M(z)$ which depend upon a spectral parameter $z$, 
and which are such that the Lax equation (2.3) is equivalent to the elliptic
Calogero-Moser Hamiltonian-Jacobi equations for arbitrary value of $z$. 
The Lax operators are not unique since the Lax equation is invariant under 
the 
following gauge transformations by an arbitrary $N\times N$ matrix-valued 
function $S$ of $x$, $p$ and $z$,
$$
\eqalign{
L \to L ^S & = S L S^{-1} \cr
M \to M ^S & = S M S^{-1} - \dot S S^{-1}. \cr}
\eqno (2.4)
$$
In particular, the action of the Weyl group $W_\G$ of $\G$ on the operators 
$L$
and $M$ is realized in terms of such a transformation. 
As a result, the spectral curve, defined in (1.1) is invariant under the 
Weyl 
group $W_\G$ and under time evolution. 

\medskip

We shall now summarize the final expressions for the Lax operator $L(z)$ 
obtained in [17].\footnote{*}{Some key facts about Lie algebra theory are 
collected in the Appendix \S A of [17]; useful general sources are in [24].} 
The form of the operator $M(z)$ will not be needed for our 
purposes, since it does not enter into the form of the curve or of the 
differential in (1.1). We choose a Cartan subalgebra $\H _\G$ of $\G$, with 
generators $h_j$, $j=1,\cdots ,n$, conveniently assembled into a vector of 
generators, denoted by $h$. The Lie algebra $\G$ is embedded into 
$GL(N,\C)$ by via an $N$-dimensional representation $\Lambda$ with weight 
vectors $\lambda _I$, $I=1,\cdots ,N$. We denote the generators of 
$GL(N,\C)$ 
by $E_{IJ}$ with $I,J=1,\cdots ,N$. The general form of $L(z)$ for both 
ordinary 
and twisted Calogero-Moser systems is then given by
$$
L (z) = p \cdot h + \sum _{I,J=1; I\not= J} ^N C_{I,J} \Phi _{IJ} 
\bigl ( (\lambda _I  - \lambda _J) \cdot x,z \bigr ) E_{IJ}.
\eqno (2.5)
$$
We shall establish in \S III below that the systems relevant to 
super-Yang-Mills 
dynamics for gauge algebra $\G$ are the ordinary (resp. twisted) 
Calogero-Moser system for simply laced $\G$ (resp. non-simply laced $\G$). 
We 
list the entries $C_{I,J}$ and $\Phi _{IJ}$ separately for each case.

\bigskip

\noindent
{\bf (a) Simply Laced $\G$ : Ordinary Calogero-Moser Systems}

\medskip

The elliptic functions $\Phi _{IJ}$ are independent of $I$ and $J$ and given 
by [17], (see also [11])
$$
\Phi _{IJ} (u,z) = \Phi (u,z) \equiv { \sigma (z-u) \over \sigma (z) \sigma 
(u)} e^{u \zeta (z)},
\eqno (2.6)
$$
where $\sigma (u)$ and $\zeta (u)$ are the standard Weierstrass 
functions\footnote{*}{A useful source for information on elliptic functions 
is 
[25].} 
satisfying $\sigma (u) = u + \O (u^5)$, $\zeta (u) = \sigma '(u) / \sigma 
(u)$ 
and $\zeta '(u) = - \wp (u)$. To define the constants $C_{I,J}$, we fix the 
representations of $\G$ to those of smallest (non-trivial) dimension. For 
$\G 
= A_n, \, D_n, \, E_6, \, E_7$ we have respectively $N=n+1,\, 2n,\, 27,\, 
56$. 
Each of these representations only has non-zero weights $\lambda$, which 
belong to a single orbit of the Weyl group $W_\G$. It is very convenient to
replace the labels $I$ and $J$ in (2.5) by the $N$ weight vectors of the
representation. The precise mapping between the labels $I$ and $\lambda$ is
immaterial, since they will be permuted under the action of the Weyl group. 
We
have for $\G  = A_n, \, D_n, \, E_6,\, E_7$ the following expressions [DP]
$$
 C_{I,J} = C_{\lambda , \mu} = \left \{ \matrix{
 m & \lambda - \mu  \in \R (\G)  \cr
 0 & \lambda - \mu  \notin \R (\G) . \cr}
 \right .
 \eqno (2.7)
$$
For $\G = E_8$, we have $N=248$. This representation contains 240 non-zero
weights $\lambda$ (which are roots) and 8 zero weights. It is convenient to
replace  the 248 labels $I$ and $J$ by the 240 nonzero weights $\lambda$ and
$\mu$ and by $a,b=1,\cdots, 8$ which label the zero weights. We have then 
for
$\G =E_8$
$$
\eqalign{
C_{a,b} & =0 \qquad \qquad \qquad a,b = 1,\cdots ,8 \cr
C_{\lambda , \mu} & = \left \{ \matrix{
 m \, c(\lambda, \mu) & \lambda - \mu  \in \R (E_8)  \cr
 0 & \lambda - \mu   \notin \R (E_8)  \cr} \right . \cr
C_{\lambda, a} & = \half m \sum _{b=1} ^8 c(\lambda, \beta _b) O_{\beta 
_b,a}.
\cr}
 \eqno (2.8)
$$
Here, $\beta _b$, $b=1,\cdots ,8$ is a choice of 8 (which is the maximum 
number) mutually orthogonal roots of $E_8$, corresponding to the maximal 
subalgebra $[A_1] ^{\times 8}$ of $E_8$. The $8\times 8$ matrix $O_{\beta
_b,  a}$ is an arbitrary orthogonal matrix. The functions $c(\lambda, \mu)$ 
take values $\pm 1$ only, and are defined by a set of equations discussed in 
[17].

\bigskip 

\noindent
{\bf (b) Non-Simply Laced $\G$ : Twisted Calogero-Moser Systems} 

\medskip

To define the functions $\Phi _{IJ}$ and the constants $C_{I,J}$, we fix the 
representations to be of smallest dimension. For non-simply laced $\G=B_n, 
\, 
C_n, \, F_4$, we have respectively $N=2n,\, 2n+2, \, 24$. For $\G=G_2$, only 
partial results on the existence of a Lax pair could be obtained in [17]; as 
a 
result, we shall refrain from discussing this case here.
Several different functions now enter into (2.5),
$$
\eqalign{
B_n \qquad & \Phi _{IJ}(x,z) = \left \{ \matrix{
\Phi (x,z) & ~ I-J \not=0,\pm n \cr
\Phi _2 (\12 x,z)  & I-J = \pm n \cr} \right . \cr
C_n \qquad & \Phi _{IJ} (x,z) = \Phi (x + \omega _{IJ},z) \cr
F_4 \qquad & \Phi _{\lambda \mu} (x,z)  = \left \{ \matrix{
\Phi (x,z) & \lambda \cdot \mu =0 \cr
\Phi _1 (x,z) & \lambda \cdot \mu = \half \cr
\Phi _2 (\12 x,z)  & \lambda \cdot \mu =-1 \cr} \right . \cr
\cr}
\eqno (2.9)
$$
The new functions are defined in terms of the fundamental function $\Phi 
(x,z)$ of (2.6) by the following relations. For more details, see [17].
$$
\eqalign{
\Phi _1 (x,z) &= \Phi (x,z) - \Phi (x+\omega _1,z) e^{\pi i \zeta (z) + z 
\zeta (\omega _1)} 
\cr
\Phi _2 (x,z) 
&= \Phi (x,z) \Phi (x+\omega _1,z) \Phi (\omega _1,z) ^{-1}. \cr}
\eqno (2.10)
$$
The constants $C_{I,J}$ are given by 
$$
\eqalignno{
B_n \qquad & C _{IJ}(x,z) = \left \{ \matrix{
m & ~ I-J \not=0,\pm n \cr
\sqrt 2 m_1  & I-J = \pm n \cr} \right . 
& (2.11a)\cr
C_n \qquad & C _{IJ} = \left \{ \matrix{
m & I,J=1,\cdots ,2n; I-J \not=\pm n \cr
\sqrt 2 m  & I=1,\cdots,2n, J=2n+1,2n+2;~ I\leftrightarrow J \cr
2m & I=2n+1, J=2n+2; I\leftrightarrow J \cr} \right .
& (2.11b) \cr
F_4 \qquad & C _{\lambda \mu}   = \left \{ \matrix{
m & \lambda \cdot \mu =0 \cr
0 & \lambda \cdot \mu = - \12 \cr
{1 \over \sqrt 2} m_1 & \lambda \cdot \mu = \half \cr
\sqrt 2 m_1 & \lambda \cdot \mu =-1 \cr} \right . 
& (2.11c) \cr}
$$
The cocycle shifts $\omega _{IJ}$ are given by
$$
\omega _{IJ} = \left \{ \matrix{
+ \omega _2 & I=1,\cdots ,2n+1; J=2n+2, \cr
- \omega _2 & J=1,\cdots ,2n+1; I=2n+2, \cr
0 & {\rm otherwise}. \cr} \right .
\eqno (2.12)
$$
We note here that the twisted Calogero-Moser systems for $B_n$ and $F_4$ 
involve two independent Calogero-Moser couplings $m$ and $m_1$. We shall 
discuss their physical significance at the end of \S III. (b).

\bigskip
\bigskip

\centerline{\bf III. CURVES, DIFFERENTIALS FROM CALOGERO-MOSER}

\bigskip

Given the Lax operators for the ordinary Calogero-Moser systems associated 
with 
simply laced Lie algebras $\G$, and of the twisted Calogero-Moser systems 
associated with non-simply laced Lie algebras $\G$, our proposal for the 
Seiberg-Witten curves $\Gamma$ and differentials is
$$
\eqalign{
\Gamma \ : \ R(k,z) = & ~ \det (k I - L(z)) =0 \cr
d\lambda = & ~ kdz. \cr}
\eqno (3.1)
$$
The quantum order parameters $a_i$, their dual $a_{Di}$, $i=1,\cdots ,n$ and
the prepotential $\F$ are then defined by
$$
a_i = {1 \over 2 \pi i} \oint _{A_i} d\lambda ,
\qquad
a_{Di} = {1 \over 2 \pi i} \oint _{B_i} d \lambda,
\qquad 
{\partial \F \over \partial a_i} = a_{Di}.
\eqno (3.2)
$$
Here, the cycles $A_i$ and $B_i$, $i=1,\cdots ,n$ are constructed below.
We shall now carry out the consistency checks, discussed in the 
introduction. 

\bigskip

\noindent
{\bf (a) Analytic behavior, Weyl Invariance, Counting Moduli}

\medskip

$\bullet$ The function $R(k,z)= \det(kI-L(z))$ is polynomial is $k$ and
meromorphic as a function of $z$, despite the fact that the entries 
$L_{IJ}(z)$
of the matrix
$L(z)$ themselves have essential singularities, as can be seen from the very
definition of $\Phi$ in (2.6). In fact, the expression for $L(z)$ in (2.5) 
shows
that conjugation of $L(z)$ by the diagonal matrix $S$ with components
$$
S_{IJ}(z)=\delta_{IJ} e^{\lambda_I\cdot x \,\zeta(z)}
\eqno (3.3)
$$ 
leads to an operator $L^S (z)$, defined by (2.4) with meromorphic entries. 
For
simply laced $\G$, this follows from the fact that only $\Phi$ enters (see
(2.6)), and from the fact that under the transformation (3.3), $p$ is
unchanged and the function $\Phi(u,z)$ is effectively replaced by the 
meromorphic
function 
$$
\tilde \Phi (u,z) = {\sigma(z-u)\over\sigma(z)\sigma(u)}.
\eqno (3.4)
$$
For non-simply laced $\G$, it follows from their definition in (2.10) that 
the
functions $\Phi_1(u,z)$, $\Phi _2 (u,z)$ have the same
essential singularity as $\Phi (u,z)$ itself. Thus, the argument presented 
for
simply laced $\G$ then also holds for non-simply laced $\G$. 

\medskip

$\bullet$ The Weyl group $W_\G$ is generated by Weyl reflections $W_\alpha$,
$\alpha \in \R (\G)$, which act on $x$, and $p$ in the standard way
$$
x \to W_\alpha (x) = x - 2 \alpha \, {x\cdot \alpha \over \alpha ^2},
\qquad \qquad
p \to W_\alpha (p).
\eqno (3.5)
$$
The action of $W_\alpha$ preserves the inner product, $W_\alpha (x) \cdot
W_\alpha (y) = x \cdot y$. Thus, a Weyl reflection on $x$ and $p$ in $L(z)$ 
may
be recast in terms of the action of the Weyl reflection on the weights
$\lambda _I$ and on the Cartan generators $h$. This action is given by
$$
\eqalign{
W_\alpha (\lambda _I) & = \sum _{J=1} ^N( S _\alpha) _{IJ} \lambda _J \cr
W_\alpha (h) & = S_\alpha  h S _\alpha ^{-1} \cr 
W_\alpha (L(z)) & = S_\alpha  L(z) S _\alpha ^{-1}, \cr}
\eqno (3.6)
$$
where $S_\alpha $ is a permutation matrix with entries $(S_\alpha )_{IJ}$ 
defined
by the first line in (3.6). Thus, the action of the Weyl group on $L(z)$ is
simply by conjugation, and the spectral curve (3.1) is invariant under 
$W_\G$.

\medskip

$\bullet$ The curves $R(k,z)=0$ are expected to depend upon precisely $n$ 
complex moduli, which are the independent integrals of motion of the 
Calogero-Moser system. Let us briefly explain why. Each Lax operator $L(z)$ 
depends upon all of the $2n$ degrees of freedom $x_i$ and $p_j$, 
$i,j=1,\cdots ,n$, with a non-degenerate Poisson bracket $\{ x_i, p_j\} = 
\delta _{ij}$. The quantities $\tr L(z)^{\gamma +1}$ are all integrals of 
motion for $\gamma =0,\cdots ,\infty$. On general grounds, at most $n$ of 
these can be functionally independent. By taking the $m\to 0$ limit, one 
establishes that precisely $n$ values of $\gamma = \gamma _i$, $i=1,\cdots 
,n$ yield functionally independent integrals of motion, with the $\gamma _i$ 
corresponding to the exponents of the Lie algebra $\G$, as given in [1], 
Appendix A, Table 4. By continuity in $m$, these integrals of motion are 
also expected to be mutually independent for $m \not=0$. Thus, we have precisely 
$n$ 
functionally independent integrals of motion for all $m$.

\medskip

Using the fact that time evolution acts by conjugation on $L(z)$, we 
immediately derive from (3.1) that the function $R(k,z)$ is conserved under 
time evolution,
$$
{d \over dt} R(k,z) = \{ H, R(k,z)\} =0.
\eqno (3.7)
$$
Thus, $R(k,z)$ must be a function of only the $n$ independent integrals of 
motion $\tr L(z)^{\gamma _i +1}$, which in super-Yang-Mills theory play the 
role of {\it moduli}, parametrizing the supersymmetric vacua of the gauge 
theory. In the case of the elliptic Calogero-Moser system for $\G = 
SU(n+1)$, this result was further confirmed in [7], where the explicit 
dependence upon the $n$ integrals of motion was exhibited explicitly, as 
will be discussed also in \S IV below.

\bigskip

\noindent
{\bf (b) Meromorphicity of $d\lambda$, pole structure}

\medskip

$\bullet$ 
Meromorphicity of the Seiberg-Witten differential $d\lambda=kdz$ is 
readily established, once it is realized that the spectral curve 
may be expressed as $R(k,z)=\det (kI - L^S (z))$, where $S$ was defined in 
(3.3) and the entries of $L^S(z)$ are meromorphic functions of $z$.

\medskip

$\bullet$ A simple pole in $z$ arises in $L ^S (z)$ when $z=0$, or more
generally when $z$ approaches $z_P= 2 \omega _1 n_1 /\nu + 2 \omega _2 n_2$, 
for $n_1,n_2 \in \Z$. The behavior at these poles is readily read off
from the behavior of $L^S(z)$, which may be derived from the
structure of $L(z)$ in (2.5) and from (3.3). We find 
$$
L ^S(z)=-{C_{I,J} \over z - z_P}+\ {\rm regular\ terms}.
\eqno(3.8)
$$
From the explicit expressions given in (2.8), (2.9) and (2.11) for 
the constants $C_{I,J}$, it is clear that 

\item{(i)} the position and the residues of the poles are
independent of the moduli, 
\item{(ii)} the residues of the poles are linear in the
hypermultiplet mass $m$.

This shows that the residues of $k$ are linear functions of $m$.
Note however that it is more difficult to determine
their exact values for general $\G$ than it was for the SU($N$) case,
when they were all $-m$, except for the last coefficient which is $(N-1)m$.

\medskip

There is one further important issue concerning the mass of the adjoint 
hypermultiplet that still needs to be addressed.
The Calogero-Moser systems for simply laced $\G$, as well the twisted 
Calogero-Moser system for $\G=C_n$ involve only a single Calogero-Moser 
coupling $m$, as in (2.7), (2.8) and (2.11b), and this parameter is 
identified 
with the mass of the hypermultiplet in the adjoint representation of $\G$ 
since 
it arises as a residue of a pole of the Seiberg-Witten differential by(3.8). 
Remarkably, the twisted Calogero-Moser systems for $B_n$ and $F_4$, (and as 
far 
as we know also $G_2$) involve two Calogero-Moser couplings $m$ and $m_1$. 

\medskip

Let us begin by discussing the case of $\G=B_n$. From considering the subset 
of 
roots of $B_n$ associated with the subalgebra $D_n$, it follows immediately 
that the coupling $m$ in (2.11a) is exactly the mass of the adjoint 
hypermultiplet. Understanding the role of $m_1$ is slightly more delicate. 
At 
the level of the integrable system, this coupling is \`a priori unrelated to 
$m$. At the level of Seiberg-Witten theory however, there can be only a 
single 
mass parameter for the adjoint hypermultiplet, since the latter transforms 
under a single irreducible representation of the gauge algebra $\G$. Given 
that 
the residue of the pole at the half period $z=\omega _1$ is linear in 
$C_{I,J}$, in view of (3.8), and takes the value $m_1$, we see that by the 
general arguments of Seiberg-Witten theory, $m_1$ must be linear in $m$, the 
hypermultiplet mass. The precise coefficient does not appear to be 
determined 
from Calogero-Moser dynamics.

\medskip

The ratio $m_1/m$ may be fixed by comparing for example the one loop 
contribution to the prepotential, obtained from standard field theory 
methods, with the result derived from the Calogero-Moser approach. It is 
likely that the special value selected this way by Seiberg-Witten theory 
corresponds to a point of enhanced symmetry at the level of the classical 
Calogero-Moser system. 
At present, we do not know what that symmetry might be, and leave this issue 
open for further investigation.

\bigskip

\noindent
{\bf (c) Structure of Homology Cycles}

\medskip

Here, we specify a set of homology cycles $A_i$ and $B_i$, $i=1,\cdots ,n$, 
for the spectral curve $\Gamma$, (where $n$ is the rank of $\G$), to be used 
in the evaluation of the quantum order parameters and effective prepotential in 
(3.2). 

\medskip

Let $A$ and $B$ be a canonical basis of homology cycles on the base 
elliptic curve $\Sigma$, and let $A$ be the cycle which shrinks to a point 
when $\tau \to +i \infty$. The spectral curve $\Gamma$ is obtained by gluing
along certain cuts $N$ copies of $\Sigma$. In the limit $m \to 0$, 
$L(z)=p\cdot h$ admits $N$ constant eigenvalues of which $n$ are 
linearly independent. Select the $n$ copies of $\Sigma$ corresponding to such
a maximal set of linearly independent eigenvalues. The desired $A_i$ and $B_i$
cycles, $i=1,\cdots ,n$ are obtained by lifting to these sheets the
$A$ and $B$ cycles of the base torus $\Sigma$.

\medskip

This prescription has been shown to reproduce the correct prepotential
in the case of $SU(N)$ in [7]. We shall show below, explicitly, that it is
also appropriate for $D_n$. See also [3] for a prescription of Prym varieties
when the spectral curve arises from a group theoretic gluing of
several copies of the sphere.

\bigskip

\noindent
{\bf (d) The limit m$ \to $0 to $\N=4$ super-Yang-Mills}

\medskip

In the $m\to 0$ limit, the effective prepotential of
the Calogero-Moser system should reproduce the classical
metric $ds^2=({\rm Im}\,\tau)\sum_{i=1}^rda_id\bar a_i$ on the space of 
vacua.
This metric is known to receive no quantum corrections, since
the $m=0$ limit is the $\N=4$ theory.

\medskip

In the original work of Donagi-Witten [4] for the SU($N$) case,
the verification of this requirement was carried out via a Hitchin system.
In terms of Calogero-Moser systems,
the verification is even simpler. Indeed, as we saw in (c) above,
at $m=0$, the Lax operator $L(z)$ reduces to $L(z)=p \cdot h$ for all 
$z\in\Sigma $, so that the spectral curve $\Gamma$, given by $\det(kI-L(z))=0$, 
reduces to $N$ {\it unglued} copies of
$\Sigma$, indexed by the constant eigenvalues of $L(z) = p \cdot h$. 
Of these only $n$ are linearly independent, say $k_1,\cdots, k_n$.
The $A_i$ and $B_i$ cycles, $i=1,\cdots, n$, are the lifts to the corresponding 
sheets of the $A$ and $B$ cycles on $\Sigma$. Thus, both the order parameters 
$a_i$ and their duals $a_{Di}$ may be evaluated in the
$m\to 0 $ limit and we find
$$
\eqalign{
a_i & 
= {1 \over 2 \pi i } \oint _{A_i} d \lambda 
= {1 \over 2 \pi i } \oint _A dz
= {2 \omega _1 \over 2 \pi i} k_i 
\cr
a_{Di} & 
= {1 \over 2 \pi i } \oint _{B_i} d \lambda 
= {1 \over 2 \pi i } \oint _B dz 
= {2 \omega _1 \over 2 \pi i} \tau  k_i 
\cr}
\eqno (3.9)
$$
 The prepotential ${\cal F}$, also defined in (3.2) is
then easily read off in the $m\to 0 $ limit and we have 
$$
{\cal F}={\tau\over 2}\sum_{i=1}^ra_i^2.
\eqno(3.10)
$$
As a result, ${\rm Im}\,\partial_{a_i} \partial _{a_j}{\cal F} = \Im \tau 
\delta
_{ij}$ correctly reproduces the classical metric.

\bigskip

\noindent
{\bf (e) The limit $m \to \infty$ to the theory without hypermultiplets}

\medskip

All the requirements analyzed above would be fulfilled by the ordinary as 
well
as by the twisted Calogero-Moser systems associated with the gauge algebra
$\G$. Thus, the final requirement on the behavior of the system as $m \to
\infty$ is crucial in distinguishing between these two possibilities, at 
least 
in the case of non-simply laced $\G$ where the twisted system differs from 
the 
ordinary one.  

\medskip

As $m\to \infty$, standard renormalization group (and $R$-symmetry) 
arguments 
dictate the dependence of the
gauge coupling $g$ and the angle $\theta$ of (1.1) on the mass $m$, in terms 
of
a renormalization scale $M$, which is kept fixed in the limit,
$$
\tau = {i \over 2 \pi} h_\G ^\vee \, \ln {m^2 \over M^2}.
\eqno (3.11)
$$
Here, $h_\G ^\vee$ is the dual Coxeter number of the gauge algebra $\G$, 
which
coincides with the quadratic Casimir of the algebra $\G$. Physically, when
$m\to \infty$ while $\tau$ obeys (3.11), the $\N=2$ super-Yang-Mills theory
with gauge algebra $\G$ and one hypermultiplet in the adjoint representation 
of
$\G$ converges to the $\N=2$ super-Yang-Mills theory {\it without 
hypermultiplets}. In order to describe the $\N=2$ super-Yang-Mills theory 
with
one adjoint hypermultiplet, the associated Calogero-Moser system 

\item{(1)} must converge to a finite limit,
\item{(2)} which must give an integrable system for the theory without 
hypermultiplets.

In a companion paper [19], we have systematically analyzed the limits of the
ordinary and twisted Calogero-Moser systems according to general scaling 
behavior of the form 
$$
m= M e^{ \pi i \delta \tau}.
\eqno (3.12)
$$
(The variable $x$ is shifted as well, but $p$ is kept fixed upon
taking the limit; we shall not need the precise form of this behavior here.)
The results from [19] relevant to this analysis are as follows. 

\item{(i)} The ordinary Calogero-Moser system has a finite limit for all
$0< \delta \leq 1/h_\G$, and diverges when $\delta > 1/h_\G$.
At the critical scaling with $\delta = 1/h_\G$, the Calogero-Moser 
Hamiltonian
and Lax pair tend to those of the (affine) Toda system associated with the
untwisted affine Lie algebra $\G ^{(1)}$.
\item{(ii)} The twisted Calogero-Moser system has a finite limit for all
$0<\delta \leq 1/h_\G ^\vee$, and diverges for $\delta >1/h_\G
^\vee$. At the critical scaling with $\delta = 1/h_\G ^\vee$, the
Calogero-Moser Hamiltonian and Lax pair tend to those of the (affine) Toda
system with the dual algebra $(\G ^{(1)})^\vee$.

\medskip

Now, the scaling (3.12) of the Calogero-Moser systems agrees with that 
required
by the renormalization group when 
$$
\delta = {1 \over h_\G ^\vee}.
\eqno (3.13)
$$
By comparing the scalings of (i) and (ii) above and the value of (3.13), we
find the following requirements on the scaling behavior.

\medskip

For {\it simply laced } $\G$, $h_\G ^\vee = h_\G$ and the twisted and 
ordinary
Calogero-Moser systems coincide. From (i), the limit is finite, confirming 
(1)
above. Furthermore, the limit is the Toda system for $\G ^{(1)}$, thus
reproducing the result of [3], as required by (2). This confirms
that for simply laced $\G$, the ordinary Calogero-Moser system indeed passes
the last consistency test of (d).

\medskip

For {\it non-simply laced } $\G$, it is always true that $h_\G > h_\G 
^\vee$.
If $\delta$ is given by (3.13), as required by the renormalization group
arguments of (3.11), then the ordinary Calogero-Moser system will diverge as
the limit $m\to \infty$ is taken according to (3.12), violating the 
requirement
(1) above. On the other hand, when $\delta$ is given by (3.13), the 
twisted Calogero-Moser system converges to the (affine) Toda system for the
affine Lie algebra $(\G ^{(1)})^\vee$. Thus, requirement (1) above is
satisfied and the limit reproduces the result of [3], as required
by (2). This confirms that for non-simply laced $\G$, the twisted
Calogero-Moser system passes the last consistency test of (d).
 
\bigskip
\bigskip

\centerline{\bf IV. CURVES FOR LOW RANK; WEAK COUPLING $D_n$ }

\bigskip

In our treatment [7] of the case $\G = SU(N)$, we succeeded in reformulating
equation (3.1) for the Seiberg-Witten curve in terms of a very simple
expression involving the Jacobi theta function $\vartheta _1$, 
$$
\vartheta _1 \bigl (
{1 \over 2 \pi i} (z - m {\partial \over \partial k})|\tau 
\bigr ) H( k) =0.
\eqno (4.1)
$$
Here, $H( k)$ stands for a polynomial of degree $N$, whose overall
normalization may be fixed to be $H( k) =  k^N + \O (k^{N-2})$. We may 
re-express $H( k) = \det ( k I- \bar k \cdot h)$, where $h$ are the Cartan
generators of $\G$ and the $N-1$ free parameters $\bar k_i$ play the role 
of the classical order parameters of the super-Yang-Mills theory. (The 
variable
$ k$ in (4.1) actually differs from that used in (3.1) via a shift by a 
function that depends upon $z$ and $\tau$ but is independent of the moduli, 
and 
is thus irrelevant for our considerations.)

\medskip

In the case of general gauge algebra $\G$, we expect the Seiberg-Witten 
curves
of (3.1) to admit simplified expressions  analogous to those for the $SU(N)$
case given in (4.1), where the role of $H( k)$ is played by $\det
 (kI - \bar k \cdot h)$. We plan to address this problem in a subsequent 
publication.
 
 \medskip
 
We now present a considerable simplification in the evaluation of the 
spectral 
curve $\Gamma$ of (3.1), by making a judicious choice of classical order 
parameters. At $m=0$, the curve $\Gamma$ is given exactly by $\det (kI - p 
\cdot h)=0$, which depends only upon $k$ and the $n$ independent Casimir 
invariants $u_i ^0$. The $u_i ^0$ are polynomials in $p$, homogeneous of 
degree 
$\gamma _i +1$, where the {\it exponents} $\gamma _i$ are given in [17], 
Appendix \S A, Table 4. At $m\not=0$, the curve $\Gamma$ depends upon both 
$p_i$ 
and $x_i$. However, the fact that the spectral curve is built on an 
integrable 
system guarantees that $ \Gamma$ depends on $p$ and $x$ only through $n$ 
combinations $u_i= u_i (m)$, which are polynomial in $m$ and satisfy $u_i 
(0) = 
u_i ^0$. Thus, $u_i(m)$ may be viewed as the deformation of $u_i^0$ away 
from 
$m=0$, and may be identified by the leading $p$ behavior. Thus, to compute 
$\Gamma$ in terms of the Casimirs $u_i(m)$, it suffices to carry out the 
calculation of the determinant for any arbitrary convenient choice of the 
variables $x_i$, since the $p$-dependence alone will allow for the 
identification of the Casimirs $u_i(m)$. One very convenient choice for $x$ 
is 
in terms of the level vector $\rho ^\vee$ of the Lie algebra $\G$, and the 
associated level function $l(\alpha)$ 
$$
x= \xi \rho ^\vee,
\qquad \qquad
\alpha \cdot x = \xi \, l(\alpha).
\eqno (4.2)
$$
Here, the parameter $\xi$ is arbitrary. Direct calculations of the curves 
$\Gamma$ are still cumbersome for large rank and for exceptional algebras. 
An 
indirect method in which the trigonometric limit (i.e. zero gauge coupling) 
is 
evaluated first allows for further simplifications, as will be explained in 
(c) 
below. 

\medskip

\noindent
{\bf (a) Curves for Low Rank Classical $\G$}

\medskip

For low rank classical groups, we have the following explicit forms of the 
spectral curves. For $\G = B_2 = C_2$, the curve reads 
$$
\eqalign{
0=& ~
k^4 -2k^2 (u_2 - 2 m ^2 \wp (z) - m_1 ^2 \wp _2 (z)) + 4 m_1 m ^2 k \wp 
'(z)
+ m_1 ^4 \wp _2 (z) ^2 \cr
& - m_1 ^2 m ^2 \wp (z) ^2 + 2 m_1 ^2 u_2 \wp _2 (z)
+ 4 m_1 ^2 m ^2 \wp _2 (z) \wp (\omega _1) + u_4.
\cr}
\eqno (4.3)
$$
Here, $u_2$ and $u_4$ are two independent classical order parameters, and 
$\wp 
_2$ is the twisted Weierstrass function, defined in (2.2). Notice that it 
may 
be expressed in terms of $\wp $ alone via the relation
$$
\wp _2 (z) = \wp (z) + { (\wp (\omega _1) - \wp (\omega _2)
(\wp (\omega _1) - \wp (\omega _3) \over \wp (z) - \wp (\omega _1) }.
\eqno (4.4)
$$
A discussion of the interpretation of the mass parameter $m_1$ was given at 
the 
end of \S III (b).

\medskip

For $\G = D_n $, the curves are
$$
0= \sum _{j=0}^n Q_{2j}(k)  u_{2n-2j},
\eqno (4.5)
$$
where $u_{2n-2j}$ are the Casimir invariants with $u_0=1$. The functions 
$Q_{2j}(k)$ are polynomials in $k$ of degree $2j$. To order $j\leq 5$, we 
have
$$
\eqalign{
Q_0 & =1 \cr
Q_2 & = k^2 \cr
Q_4 & = k^4 - 4 k^2 m^2 \wp \cr
Q_6 & = k^6 -12 k^4 m^2 \wp - 8 k^3 m^3 \wp ' \cr
Q_8 & = k^8 -24 k^6 m^2 \wp -32 k^5 m^3 \wp ' -48 k^4 m^4 \wp ^2 +64 g_2 k^2 
m^6 \wp \cr
Q_{10} & = k^{10} -40 k^8 m^2 \wp -80 k^7 m^3 \wp ' -240 k^6 m^4 \wp ^2 - 64 
k^5 m^5 \wp \wp ' \cr
& \ + 704 g_2 k^4 m^6 \wp + 512 g_2 k^3 m^7 \wp' -768 k^2 m^8 g_3 \wp, \cr}
\eqno (4.6)
$$
where we have used the abbreviations $\wp = \wp (z)$, $\wp' = \wp' (z)$ and 
where $g_2$ and $g_3$ stand for the modular forms of degrees 4 and 6 
respectively, normalized by the equation $\wp '^2 = 4 \wp ^3 - g_2 \wp 
-g_3$.
The combination of (4.5) and (4.6) yields the $D_n$ curves for $n=2,3,4,5$. 
It would not be easy to establish these low order curves by direct 
calculation of the determinants in (1.2), even using the simplifications 
explained in the preceding paragraph. Instead, indirect methods, developed 
in (c) below, were used to derive (4.5) and (4.6). We expect that a more 
general method can be found, analogous to the one used for $\G=SU(N)$ to 
derive (4.1), from which (4.6) may be obtained for general $D_n$.

\medskip

\noindent
{\bf (b) Trigonometric Calogero-Moser and Perturbative Limit : $D_n$ 
Example}

\medskip

The perturbative limit of gauge theory corresponds to $g \to 0$ in (1.1), 
which 
implies that $\tau \to +i \infty$ and $q \to 0$. One further confirmation 
that 
we have indeed uncovered the correct curves for general Lie algebras $\G$ is 
that the correct perturbative limit for the prepotential will be reproduced 
by 
these curves. For the sake of brevity, we discuss only the case $\G = D_n$ 
here.

\medskip

At the level of the elliptic Calogero-Moser system, the perturbative limit 
produces the trigonometric Calogero-Moser system in which 
$$ 
\eqalign{
\wp (z) \ \to & \ { 1 \over Z^2} - {1 \over 6} 
= {1 \over 4} { 1 \over \sinh ^2 {z \over 2} } + {1 \over 12}\cr
\Phi (x,z) \ \to & \ \half \coth \half x - {1 \over Z} \cr
{1 \over Z} \ = & \ \half \coth \half z .\cr}
\eqno (4.7)
$$
Here, we have introduced a natural variable $Z$ which will prove to be 
convenient shortly. The curves in the trigonometric limit may be evaluated 
completely explicitly for $\G=D_n$, and from this information, the low order 
curves of (4.5) and (4.6) may be inferred. We begin by deriving the curves 
in 
this limit. We make use of the additional simplification by choosing $x$ as 
in 
(4.3) and letting $\xi$ be real and $\xi \to + \infty$. The function 
$\Phi (\alpha \cdot x,z)$ which enters the Lax operator $L(z)$ has a 
particularly simple form in this limit, given by
$$
\Phi (\alpha \cdot x,z) \to - {1 \over Z} + \left \{
\matrix{ + \half & \alpha >0 \cr
          &\cr
         - \half & \alpha <0.  \cr} \right .
         \eqno (4.8)
$$
Introducing the $n\times n$ matrices $\mu ^\pm$ by
$$
\mu ^+ _{ij} = \left \{ \matrix{1 & i<j \cr
                                0 & i\geq j \cr} \right .
                        \qquad \qquad
\mu ^- _{ij} = \left \{ \matrix{ 1 & i>j \cr
                        0 & i\leq j, \cr} \right .
\eqno (4.9)
$$
the matrix $\mu = \mu ^+ + \mu ^-$, and $P={\rm diag} (p_1 , \cdots , p_n)$, 
the 
equation (1.2) for the curve becomes $R(k,z)=0$ with
$$
R(k,z)= \det \left ( \matrix{
kI -P +{m\over Z} \mu -{m\over 2}(\mu ^+ - \mu ^-)
&
({m \over Z}  - {m \over 2})\mu  \cr  
& \cr 
({m \over Z}  + {m \over 2})\mu  
&
kI +P +{m\over Z} \mu +{m\over 2}(\mu ^+ - \mu ^-) \cr}
\right ).
\eqno (4.10)
$$
By taking suitable linear combinations of rows and columns, one easily shows
that the evaluation of the above determinant can be reduced to the 
evaluation
of a determinant of an $n\times n$ matrix, given as follows
$$
R(k,z) = \det \bigl [
(kI +P - m \mu ^-)(kI -P -m\mu ^+) + k(m + 2 {m \over Z}) \mu  \bigr ].
\eqno (4.11)
$$        
The determinants of the factors $kI\pm P - m \mu ^\mp$ in the first term in 
the 
brackets are easy to compute since each factor is a triangular matrix. 
However, 
the second term in the brackets would seem to spoil this advantage. 
Actually, 
by rearranging the expansion of both terms in the bracket, the determinant 
can 
be expressed as follows
$$
R(k,z) = \det \bigl [
(AI +P - m \mu ^-)(AI -P -m\mu ^+) + (mA + 2k {m \over Z}) (\mu +I) \bigr ],
\eqno (4.12)
$$        
where the new variable $A$ is defined by a quadratic relation in terms of 
$k$ 
and $Z$,
$$
0  = A^2 +mA + 2k { m \over Z} - k^2 .
\eqno (4.13)
$$
The definition is chosen in such a way that a matrix of rank 1 appears in 
the 
second term in the bracket in (4.12).
(This remarkable relation is the analogue for $D_n$ of a linear change of 
variables made for the case of $A_n$ in (3.5) of [7]; in both cases, this 
change of variables is the key relation that allows for a completely 
explicit 
solution.) Any symmetric matrix of rank 1, such as $I+\mu$ may be written in 
terms of a column vector $u$ as $I+\mu = u u^T$.
We use this fact and the following fundamental relation 
$$
\det [M + uu^T] = \det M ( 1 + u^T M^{-1} u),
\eqno (4.14)
$$
where $M$ is any invertible matrix and $u$ is any column vector, to 
complete the evaluation of the determinant of (4.12). We find
$$
R(k,z) = \prod _{j=1} ^n (A^2 - p_j^2)
+ (mA + 2 {m \over Z}) \sum _{j=1} ^n \prod _{i=1} ^{j-1} \bigl ( (A+m)^2 - 
p_i^2 \bigr ) \prod _{i=j+1} ^n (A^2 - p_i^2).
\eqno (4.15)
$$
Further algebraic manipulations permit us to recast the final result in the 
following form,
$$
\eqalignno{
R(k,z) & = {m^2 +mA -2k {m \over Z} \over m^2 +2mA} H(A) +
{mA +2k {m \over Z} \over m^2 +2mA} H(A+m) & (4.16a) \cr
H(A) & = \prod _{j=1} ^n (A^2 - p_j ^2) = \sum _{j=0} ^n (-1)^{n-j} A^{2j}
u_{2n-2j}  & (4.16b) \cr}
$$
Here, we have identified the invariant polynomials in $p_j$ with the 
integral 
invariants of the system $u_{2n-2j}$, which are also the classical order 
parameters, as discussed in (b). The curve $\Gamma$ given by $R(k,z)=0$, in 
the 
trigonometric limit, is now simply expressed in the variables $A$ and $Z$ by
$$
(m^2 +mA -2k {m \over Z} ) H(A) + (mA + 2 k {m \over Z}) H(A+m)=0.
\eqno (4.17)
$$
This equation is remarkably close to the equation found within the same 
approximation for $A_n$ in (4.5) of [7] with $q=0$.

\bigskip

\noindent
{\bf (c) Inferring Elliptic from Trigonometric Curves for $D_n$}

\medskip

Given a spectral curve $\Gamma$ for the elliptic Calogero-Moser system (say 
for 
$\G=D_n$), the limit $q\to 0$ will produce the curves of the trigonometric 
Calogero-Moser system (up to redefinitions of the classical order 
parameters, 
which is physically irrelevant). By identifying the limit with the 
corresponding curve of (4.16), we can learn a great deal about the elliptic 
case. What will be missed are quantities that are proportional to variables 
that vanish in this limit. Clearly, $p_j$, $\wp$, $g_2$ and $g_3$ all have 
non-zero limits. However, the discriminant $\Delta$ of the underlying 
elliptic 
curve tends to zero, 
$$
\Delta = g_2 ^3 - 27 g_3 ^2 \ \to \ 0,
\eqno (4.18)
$$
since $g_2 \to 1/12$ and $g_3 \to - 1/216$. Thus, the form of the curves in 
the  
trigonometric case determines the form of the curves in the elliptic case, 
up 
to functions that vanish with $\Delta$. Since $\Delta$ enters polynomially 
in 
$R(k,z)$, and the scaling degree of $\Delta$ is 12, curves will be uniquely 
determined by their trigonometric limits when $n \leq 5$. 

\medskip

To work out the trigonometric curves from (4.16) we proceed as follows.
The combination $A$ is somewhat inconvenient, since it is not rational in 
$k$. 
Using (4.16a), we may however obtain a recursion relation for the 
coefficients 
$P_{2j}$ of the expansion 
$$
R(k,z) = \sum _{j=0} ^n (-1)^{n-j} P_{2j} u_{2n-2j}
\eqno (4.19)
$$
in terms of classical order papameters $u_{2n-2j}$ which is manifestly 
polynomial in $k$. The result is
$$
0=P_{2(j+1)} - (2k^2 + m^2 -4k {m \over Z}) P_{2j} + k^2 (k-2{m \over Z})^2 
P_{2(j-1)},
\eqno (4.20a)
$$
with the initial conditions $P_0=1,\ P_2 = k^2$. The lowest non-trivial 
orders 
are then
$$
\eqalign{
P_4  & = k^4 - 4 k^2 {m^2 \over Z^2} + m^2 k^2 \cr
P_6  & = k^6 -12 k^4 {m^2 \over Z^2} +16 k^3 {m^3 \over Z^3} - 4 k^3 {m^3 
\over 
Z} -4 k^2 {m^4 \over Z^2} + 3k^4 m^2 + k^2 m^4 \cr
P_8  & = k^8 -24 k^6 {m^2 \over Z^2} +6k^6m^2 + 64 k^5 {m^3 \over Z^3}
-16 k^5 {m^3 \over Z} -48 k^4 {m^4 \over Z^4} 
\cr
& \ -8 k^4 {m^4 \over Z^2} + 5 k^4 m^4 +32 k^3 {m^5 \over Z^3} -8k^3 {m^5 
\over 
Z} - 4 k^2 {m^6 \over Z^2} +k^2 m^6. \cr}
\eqno (4.20b)
$$
and 
$$
\eqalign{
P_{10} &= k^{10} -40 k^8 {m^2 \over Z^2} +160 k^7 {m^3 \over Z^3} -240 k^6 
{m^4 
\over Z^4} +128 k^5 {m^5 \over Z^5} \cr
& \ +m^2 [ 10 k^8 -40 k^7 {m \over Z} +160 k^5 {m^3 \over Z^3} -160 k^4 {m^4 
\over Z^4} ] \cr
& \ +m^4 [15 k^6 -48 k^5 {m\over Z} +12 k^4 {m^2 \over Z^2} +48 k^3 {m^3 
\over 
Z^3}] \cr
& \ +m^6 [7 k^4 -12 k^3 {m\over Z} -4 k^2 {m^2\over Z^2} ] + m^8 k^2. \cr
}
\eqno (4.20c)
$$
Now using the limit of $\wp$ and its derivative, as in (4.7), we may 
uniquely 
identify which functional dependence in $\wp$ gave rise to each of the terms 
in 
(4.20). Doing so, (and allowing for redefinitions of the classical order 
parameters $u_{2n-2j}$), we find the results of (4.5) and (4.6), with 
$Q_{2j} 
\to P_{2j}$.

\bigskip

\noindent
{\bf (d) Agreement with Perturbation Theory : $D_n$ Example}

\medskip

Since we now possess the Calogero-Moser curve for $\G=D_n$ in the 
trigonometric 
limit, we should be able to compute the contribution to the effective 
prepotential of the $D_n$ theory with a massive adjoint hypermultiplet to 
perturbative order. To do so, it is convenient to make use of the form 
(4.16a) 
of the curve : $R(k,z)=0$ implies the following expression for the curve
$$
e^u ={ H(A+m) \over H(A)},
\eqno (4.21)
$$
where we define the complex variable $u$ by
$$
e^u \equiv {(k+A+m)(k-A-m) \over (k+A)(k-A)}.
\eqno (4.22)
$$
Recall that $A$ was defined in (4.13) as a function of $k$ and $Z$. To 
evaluate 
the prepotential, we need the Seiberg-Witten differential $d\lambda = k dz$ 
in 
terms of the new variable $A$. This is achieved by first changing variables 
from $(k,z)$ to $(A,u)$, using (4.13) and (4.22), i.e. without using the 
curve 
equation $R(k,z)=0$. First, we obtain $z$ as a function of $Z$, by inverting 
the last line in (4.7), and then use (4.13) to express $Z$ as a function of 
$A$ 
and $k$,
$$
e^z = { +1 + {2 \over Z} \over -1 +{2 \over Z}} 
= {(k-A) (k+A+m) \over (k+A)(k-A-m)}.
\eqno (4.23)
$$
Finally, $k$ may be expressed in terms of $A$ and $u$ using (4.22). Now, it 
is 
easy to work out $d\lambda =kdz$,
$$
\eqalign{
d\lambda  & = \{ {k \over k-A} - {k \over k-A-m} \} (dk-dA)
+ \{ {k \over k+A+m} - {k \over k+A} \} (dk +dA)\cr
& = \{ {A \over k-A} - {A+m \over k-A-m} \} (dk-dA)
- \{ {A+m \over k+A+m} - {A \over k+A} \} (dk +dA),\cr}
\eqno (4.24)
$$
which is readily re-expressed in terms of $A$ and $u$,
$$
d\lambda = -A du - m d \log (k^2 - (A+m)^2). 
\eqno (4.25)
$$
The last term in (4.25), integrated around any closed curve, as is always 
the 
case in Seiberg-Witten theory, gives rise to moduli independent 
contributions 
only and is physically irrelevant. Remarkably, the curve (4.21) and the 
Seiberg-Witten differential (4.25) in terms of the variables $A$ and $u$ are 
identical to the ones for $\G=SU(N)$ in terms of the 
variables $k$ and $z$. (See [7], eq. (4.13).) Thus, the calculation of the 
effective prepotential for $D_n$ to perturbative order follows directly from 
the our calculation for $SU(N)$. We find
$$
\F ^{\rm pert} = - { 1 \over 8 \pi i}
\sum _{\alpha \in \R (D_n)} \{ (\alpha \cdot a)^2 \log (\alpha \cdot a)^2
- (\alpha \cdot a + m )^2 \log (\alpha \cdot a +m )^2 \},
\eqno (4.26)
$$ 
which agrees with the standard perturbation theory result for the effective 
prepotential of a theory with an adjoint hypermultiplet with mass $m$.

\bigskip
\bigskip

\centerline{\bf V. FURTHER RESULTS AND ISSUES}

\bigskip

\noindent
{\bf (a) The Effective Prepotential Equation}

\medskip

In the analysis of the $\N=2$ super-Yang-Mills theory with hypermultiplets 
in
the fundamental representation of (classical) gauge algebras [13] or in that
of the theory with one hypermultiplet in the adjoint representation of the
$SU(N)$ gauge algebra [7] powerful renormalization group type equation were
obtained for the prepotential. We propose that the same relation should hold
between the prepotential of the gauge theory and the Hamiltonian of the 
integrable system,
$$
\eqalignno{
a_i & = { 1 \over 2 \pi i} \oint _{A_i} dz \, k & (5.1a) \cr
{\partial \F \over \partial \tau}  
= H  & = {1 \over 4 \pi i} \oint _A dz \, \tr L^2. & (5.1b) \cr}
$$
Here, $H$ is the (twisted) elliptic Calogero-Moser Hamiltonian of (2.1). 
$H$ may be expressed solely in terms of the quantum order parameters $a_i$
and the modulus $\tau$, by inverting the relation (5.1a) to obtain $a_i$ as 
a function of the classical order parameters and $\tau$. 

\bigskip

\noindent
{\bf (f) Duality Properties}

\medskip

We consider transformations of the (half) periods $\omega _1$ and $\omega 
_2$ 
of the following form
$$
\left ( \matrix{\omega _2 \cr \omega _1} \right )
\to
\left ( \matrix{\omega _2' \cr \omega _1 '} \right )
=
\left ( \matrix{a & b \cr c & d} \right )
\left ( \matrix{\omega _2 \cr \omega _1} \right )
\qquad \qquad 
a,b,c,d \in \Z,
\eqno (5.2)
$$
with $\delta = ad-bc\not=0$. When $\delta =1$, these transformations form 
the modular group, or a subgroup thereof. For $\delta =2$ these are the 
Landen or Gauss transformations familiar from the theory of elliptic 
functions, and 
associated with mapping the period lattice into a lattice where one of the 
periods is reduced to half, while the other is left intact [25].
As we have defined it here, the spectral parameter $z$ is unchanged under
modular transformations. The Weierstrass functions $\sigma (z)$, $\zeta (z)$
and $\wp(z)$, and thus the function $\Phi (x,z)$ are similarly seen to be
invariant. 

\medskip

We immediately conclude that {\it the curves $R(k,z)=0$ for simply laced 
$\G$ are modular invariant}. Physically, for these gauge algebras, the 
super-Yang-Mills theories are thus {\it self-dual}, namely they are 
invariant under the interchange of weak and strong gauge coupling $\tau$, 
defined in (1.1), under the modular transformation $S$ by $\tau \to 
-1/\tau$. As such, these $\N=2$ super-Yang-Mills theories provide explicit 
realizations of the Montonen-Olive duality conjecture [26], just as the 
massless $\N=4$ theory does. (See also [1], [4] and [27].)

\medskip

For non-simply laced $\G$, functions other than $\sigma (z)$, $\zeta (z)$ 
and
$\wp (z)$ are involved in the expressions for the curves $R(k,z)=0$.
Specifically, non-simply laced $\G$ corresponds to twisted elliptic 
Calogero-Moser systems in which the short roots are twisted with a preferred 
half period $\omega _a$, $a=1,2,3$, taken to be $\omega _1$ here. (resp. 
third period in the case of $\G=G_2$) Having singled out a preferred half 
(resp. third) period, the full modular invariance 
is broken to a subgroup which leaves the preferred half (resp. third) period 
invariant. Defining the congruence subgroups in the usual manner,
$$
\Gamma _0(\nu) = \biggl \{ \left ( \matrix{a & b \cr c & d} \right ); \ 
ad-bc 
=1, \ 
c\equiv 0 \ ({\rm mod}\ \nu) \biggr \},
\eqno (5.3)
$$
we see that the remaining subgroup of the modular group is $\Gamma _0 (2)$ 
for 
$\G=B_n, C_n, F_4$ and $\Gamma _0 (3)$ for $\G = G_2$.

\medskip

In [17], it was shown that under one of the {\it Landen or Gauss 
transformations} [25] with $\delta =2$, the elliptic Calogero-Moser 
Hamiltonian 
for $\G = B_n, C_n, F_4$ are mapped into Calogero-Moser Hamiltonians for the 
dual algebras $\G  ^\vee  = C_n, B_n, F_4$, and that for $\delta =3$, $G_2$ 
is 
mapped into itself.  We have not been able to show anything 
analogous for the Lax operators or for the spectral curves. 
We do not know at present what the precise role of these transformations 
with 
$\delta \not=1$ is. The mapping between the Calogero-Moser Hamiltonians 
leads us 
to speculate that there may exist an underlying such symmetry of the 
spectral 
curve and perhaps of the gauge theory as well. 

\bigskip

\noindent
{\bf (c) Decoupling to smaller representations}

\medskip

Decoupling of all or part of the adjoint hypermultiplet by tuning the vacuum 
expectation values of the gauge scalar and of the hypermultiplet mass was 
used 
in [7] as a powerful tool to obtain $\N=2$ supersymmetric Yang-Mills 
theories 
with different gauge groups and with hypermultiplets in different  
representations of the gauge group. In these decouplings, we showed that 
only 
the most asympotically free part of the gauge group will survive in the 
decoupling limit (i.e. at energies low compared to the decoupling scale).
In particular, any $U(1)$ factors that may arise in the group theoretic 
decomposition of the gauge group, will not survive in the physical low 
energy 
theory, a fact also familiar from [16].

\medskip

Specifically, starting with a hypermultiplet in the adjoint representation 
of 
the gauge group $SU(N_c +N_f)$, we were able to reach, by decoupling, a 
theory 
with gauge group $SU(N_c)$ and $N_f$ hypermultiplets (with $N_f < 2N_c -1$) 
in 
the fundamental representation of $SU(N_c)$ of arbitrary masses. For certain 
special arrangements of the groups and couplings, we could also achieve 
product 
gauge groups $SU(N_1) \times \cdots \times SU(N_p)$ with hypermultiplets in 
fundamental and bi-fundamental representations, such as those solved in 
[16].

\medskip

It should be clear that the same decoupling techniques may be applied to the 
$\N=2$ super Yang-Mills theories with adjoint hypermultiplet for which we 
have 
derived curves for general gauge groups in this paper. We shall leave a 
detailed discussion for a later publication, and limit ourselves here to 
pointing out some interesting cases. 

\medskip

\item{(1)} Decoupling of $SO(2n)$ (resp. $SO(2n+1)$) to a subgroup $SU(p)$ 
with 
$1<p<n$, minimally embedded into the maximal $SU(n)$ subgroup of $SO(2n)$ 
(resp. $SO(2n+1)$) should yield a theory with $SU(p)$ gauge group and 
hypermultiplets in the fundamental and {\it rank 2 anti-symmetric} 
reps of $SU(p)$. \footnote{*}{Curves for $SU(p)$ gauge group and hypermultiplets 
in (anti-)symmetric representations of the gauge group were constructed using 
brane technology and M-theory by Landsteiner and Lopez in [22]. Explicit checks 
of perturbative contributions, and instanton corrections were carried out in 
[29].}

\item{(2)} Decoupling of $Sp(2n)$ to a subgroup $SU(p)$ with $1<p<n$, 
minimally 
embedded into the maximal $SU(n)$ subgroup of $Sp(2n)$ should yield a theory 
with $SU(p)$ gauge group and hypermultiplets in the fundamental and {\it 
rank 2 symmetric} reps of $SU(p)$. 

\item{(3)} Decoupling of $E_8$, $E_7$ or $E_6$ to one of its exceptional 
subgroups (say $E_7$ or $E_6$) is expected to yield a theory with 
exceptional 
gauge group and one or more hypermultiplets in the 56-dimensional 
rep of $E_7$ and the 27-dimensional representation of $E_6$. 

\item{(4)}  Decoupling of $E_8$, $E_7$ or $E_6$ to one of its $SO(p)$ 
subgroups 
is expected to yield a theory with $SO(p)$ gauge group and one or more 
hypermultiplets in fundamental and {\it spinor representations} of $SO(p)$.

\bigskip
\bigskip

\centerline
{\bf ACKNOWLEDGEMENTS}

\bigskip

We have benefited from useful conversations with Elena Caceres, Ron Donagi 
and Igor Krichever. The first author wishes to thank Edward Witten for a 
generous invitation to the Princeton Institute for Advanced Study, where this 
research was initiated, as well as the Aspen Center for Physics. 
He would also like to acknowledge David Gross and the members of the Institute 
for Theoretical Physics in Santa Barbara for the hospitality extended to him 
while most of this work was being carried out.
Both authors would like to thank David Morrison, I.M. Singer and Edward Witten 
for inviting them to participate in the 1998 workshop on ``Geometry and 
Duality", at the Institute for Theoretical Physics. 

\bigskip
\bigskip

\centerline
{\bf REFERENCES}

\bigskip

\item{[1]} Seiberg, N. and E. Witten, ``Electro-magnetic duality,
monopole condensation, and confinement in N=2 supersymmetric Yang-Mills
theory", Nucl. Phys. B 426 (1994) 19, hep-th/9407087;
Seiberg, N. and E. Witten, ``Monopoles, duality,
and chiral symmetry breaking in N=2 supersymmetric QCD",
Nucl. Phys. {\bf B431} (1994) 494, hep-th/9410167.

\item{[2]} Gorski, A., I.M. Krichever, A. Marshakov, A. Mironov, A.
Morozov, ``Integrability and Seiberg-Witten Exact Solution", Phys. Lett. 
{\bf 
B355} (1995) 466, hep-th/9505035; 
\next
Matone, M., ``Instantons and Recursion Relations in N=2 SUSY Gauge 
Theories", 
Phys. Lett. {\bf  B357} (1996) 342, hep-th/9506102;
\next
Nakatsu, T. and K. Takasaki, ``Whitham-Toda Hierarchy and N=2 Supersymmetric 
Yang-Mills Theory", Mod. Phys. Lett. {\bf A 11}
(1996) 157-168, hep-th/9509162; ``Isomonodromic Deformations and 
Supersymmetric 
Gauge Theories", Int. J. Mod. Phys. {\bf A11} (1996) 5505, hep-th/9603069.

\item{[3]} Martinec, E. and Warner, N., ``Integrable systems and
supersymmetric gauge theories", Nucl. Phys. {\bf B459} (1996) 97-112,
hep-th/9509161.

\item{[4]} Donagi, R. and E. Witten, ``Supersymmetric Yang-Mills
theory and integrable systems", Nucl. Phys. {\bf B460} (1996) 299-334,
hep-th/9510101.

\item{[5]} Martinec, E., ``Integrable structures in supersymmetric
gauge and string theory", Phys. Lett. {\bf B367} (1996) 91, hep-th/9510204.

\item{[6]}
Sonnenschein, J., S. Theisen, and S. Yankielowicz, ``On the Relation between 
the Holomorphic Prepotential and the Quantum Moduli in SUSY Gauge Theories", 
Phys. Lett. {\bf B367} (1996) 145-150, hep-th/9510129.
\next
Eguchi, T. and S.K. Yang, ``Prepotentials
of N=2 supersymmetric gauge theories and soliton
equations", hep-th/9510183;
\next
Itoyama, H. and A. Morozov, ``Prepotential and the Seiberg-Witten theory", 
Nucl. Phys. {\bf B491} (1997) 529, hep-th/9512161; ``Integrability and 
Seiberg-Witten theory", hep-th/9601168; ``Integrability and Seiberg-Witten 
Theory : Curves and Periods", Nucl. Phys. {\bf B477} (1996) 855, 
hep-th/9511126;
\next
Ahn, C. and S. Nam, ``Integrable Structure in Supersymmetric Gauge Theories 
with Massive Hypermultiplets", Phys. Lett. {\bf B387} (1996) 304, 
hep-th/9603028;
\next
Krichever, I.M. and D.H. Phong, ``On the integrable geometry of 
soliton equations and N=2
supersymmetric gauge theories", J. Differential Geometry {\bf 45} (1997)
349-389, hep-th/9604199;
\next
Bonelli, G., M. Matone, ``Nonperturbative Relations in N=2 SUSY Yang-Milss 
WDVV 
Equation", Phys. Rev. Lett. {\bf 77} (1996) 4712, hep-th/9605090;
\next
Marshakov, A., A. Mironov, and A. Morozov,
``WDVV-like equations in N=2 SUSY Yang-Mills theory", Phys. Lett. {\bf B389} 
(1996) 43, hep-th/9607109;
\next
Marshakov, A. ``Non-perturbative quantum theories and integrable equations",
Int. J. Mod. Phys. {\bf A12} (1997) 1607, hep-th/9610242; 
\next
Nam, S. ``Integrable Models, Susy Gauge Theories and String Theory", Int. J. 
Mod. Phys. {\bf A12} (1997) 1243, hep-th/9612134;
\next
Marshakov, A., A. Mironov, A. Morozov, "More Evidence for the WDVV
Equations in N=2 SUSY Yang-Mills Theories", hep-th/9701123;
\next
Marshakov, A., ``On Integrable Systems and Supersymmetric Gauge Theories",
Theor. Math. Phys. {\bf 112} (1997) 791, hep-th/9702083;
\next
Krichever, I.M. and D.H. Phong, ``Symplectic forms in the
theory of solitons", hep-th/9708170,
to appear in Surveys in Differential Geometry, Vol. {\bf III}.

\item{[7]} D'Hoker, E. and D.H. Phong, ``Calogero-Moser Systems in SU(N)
Seiberg-Witten Theory", Nucl. Phys. {\bf B513} (1998) 405, hep-th/9709053.

\item{[8]} Lerche, W., ``Introduction to Seiberg-Witten theory
and its stringy origin", Proceedings
of the {\it Spring School and Workshop in String Theory},
ICTP, Trieste (1996), hep-th/9611190; Nucl. Phys. Proc. Suppl. {\bf 55B} 
(1997) 
83, and references therein;
\next
Donagi, R., ``Seiberg-Witten Integrable Systems", alg-geom/9705010;
\next
Freed, D., ``Special K\"ahler Manifolds", hep-th/9712042;
\next
Carroll, R., ``Prepotentials and Riemann Surfaces'', hep-th/9802130.

\item{[9]} Hitchin, N., ``Stable bundles and integrable systems",
Duke Math. J. {\bf 54} (1987) 91.

\item{[10]} Calogero, F.,
``Solution of the one-dimensional N-body problem with quadratic
and/or inversely quadratic pair potentials", J. Math. Physics {\bf 12} 
(1971)
419-436;
\hfill\break
Moser, J., ``Integrable systems of non-linear evolution equations",
in {\it Dynamical Systems, Theory and Applications},
J. Moser, ed., Lecture Notes in Physics 38 (1975) Springer-Verlag.

\item{[11]} Krichever, I.M., ``Elliptic solutions of the
Kadomtsev-Petviashvili equation and integrable systems
of particles", Funct. Anal. Appl. {\bf 14} (1980) 282-290.

\item{[12]} M.A. Olshanetsky and A.M. Perelomov, ``Classical integrable 
finite-dimensional systems related to Lie algebras'', Phys. Rep. {\bf 71C} 
(1981) 313-400.
\hfil\break
Perelomov, A.M., {\it ``Integrable Systems of Classical Mechanics
and Lie Algebras"}, Vol. I, Birkh\"auser  (1990), Boston; 
and references therein.
\hfil\break
Leznov, A.N. and M.V. Saveliev, {\it Group Theoretic Methods for Integration 
of Non-Linear Dynamical Systems}, Birkhauser 1992.

\item{[13]} D'Hoker, E., I.M. Krichever, and D.H. Phong,
``The renormalization group equation for N=2 supersymmetric
gauge theories", Nucl. Phys. {\bf B 494} (1997), 89-104, hep-th/9610156.

\item{[14]} 
Argyres, P.C., and A. E. Faraggi, ``The Vacuum Structure and Spectrum of N=2 
Supersymmetric SU(N) Gauge Theories", Phys. Rev. Lett. {\bf 74} (1995) 3931, 
hep-th/9411057;
\next
Klemm, A., W. Lerche, S. Yankielowicz, S. Theissen, ``Simple Singularities 
and 
N=2 Supersymmetric Yang-Mills Theory", Phys. Lett. {\bf B344} (1995) 169, 
hep-th/9411048;
\next
Klemm, A., W. Lerche, and S. Theisen, ``Non-perturbative
actions of N=2 supersymmetric gauge theories",
Int. J. Mod. Phys. {\bf  A11} (1996) 1929-1974, hep-th/9505150.

\item{[15]} Hanany, A., ``On the Quantum Moduli Space of Vacua of N=2 
Supersymmetric SU(N) Gauge Theories", Nucl. Phys. {\bf B452} (1995) 283, 
hep-th/9505075;
\next
D'Hoker, E., I.M. Krichever, and D.H. Phong,
``The effective prepotential for N=2 supersymmetric
SU($N_c$) gauge theories", Nucl. Phys. {\bf B489} (1997) 179-210,
hep-th/9609041; 
``The effective prepotential
for N=2 supersymmetric SO($N_c$) and Sp($N_c$) gauge theories",
Nucl. Phys. {\bf B489} (1997) 211-222, hep-th/9609145;
\next
D'Hoker, E. and D.H. Phong,
``Strong coupling expansions in $SU(N)$ Seiberg-Witten theory",
hep-th/9701151, Phys. Lett. {\bf B397} (1997) 94-103. 

\item{[16]} Witten, E., ``Solutions of four-dimensional
field theories via M-theory", Nucl. Phys. {\bf B500} (1997) 3, 
hep-th/9703166.

\item{[17]} D'Hoker, E., and D.H. Phong, ``Calogero-Moser Lax Pairs with 
Spectral Parameter for General Lie Algebras", April 1998, hep-th/9804124.

\item{[18]} V.I. Inozemtsev, ``Lax representation with spectral parameter on 
a 
torus for integrable particle systems'', Lett. Math. Phys. {\bf 17} (1989) 
11-17.

\item{[19]} D'Hoker, D.H. and D.H. Phong, ``Calogero-Moser and Toda systems
for twisted and untwisted affine Lie algebras", April 1998 preprint, 
hep-th/9804125.

\item{[20]} Kachru, S., C. Vafa, ``Exact Results for N=2 Compactifications 
of 
Heterotic Strings", Nucl. Phys. {\bf B450} (1995) 69, hep-th/9505105;
\next
Bershadsky, M., K. Intrilligator, S. Kachru, D.R. Morrison, V. Sadov and C. 
Vafa, ``Geometric Singularities and Enhanced Gauge Symmetries", Nucl. Phys. 
{\bf B481} (1996) 215, hep-th/9605200;
\next
Katz, S., A. Klemm, C. Vafa, ``Geometric Engineering of Quantum Field 
Theories", 
Nucl. Phys. {\bf B497} (1997) 173, hep-th/9609239;
\next
Katz, S., P. Mayr, and C. Vafa, ``Mirror symmetry
and exact solutions of 4D N=2 gauge theories", Adv. Theor. Math. Phys. {\bf 
1} 
(1998) 53, hep-th/9706110.

\item{[21]} Hanany, A., and E. Witten, ``Type IIB Superstrings, BPS 
Monopoles, 
and Three-Dimensional Gauge Dynamics", Nucl. Phys. {\bf B492} (1997) 152.
 
\item{[22]} Brandhuber, A., J. Sonnenschein, S. Theisen and S. Yankielowicz, 
``M-Theory and Seiberg-Witten Curves : Orthogonal and Symplectic Groups",
Nucl. Phys. {\bf B504} (1997) 175, hep-th/9705232.
\next
Landsteiner, K., E. Lopez, ``New Curves From Branes", hep-th/9708118;
\next
Landsteiner, K., E. Lopez, DA. Lowe, ``N=2 Supersymmetric Gauge Theories,
Branes and Orientifolds", Nucl. Phys. {\bf B507} (1997) 197, hep-th/9705199;
\next
Uranga, A.M., ``Towards Mass Deformed N=4 SO(N) and Sp(K) Gauge Theories 
from 
Brane Configurations", hep-th/9803054;
\next
Yokono, T., ``Orientifold four plane in brane configurations and N=4 
USp(2N) and SO(2N) theory'', hep-th/9803123.

\item{[23]}
Gorskii, A., ``Branes and Integrability in the N=2 SUSY YM Theory", Int. J. 
Mod. Phys. {\bf A12} (1997) 1243, hep-th/9612238;
\next
Gorskii, A., S. Gukov and A. Mironov, ``Susy Field Theories, Integrable 
Systems 
and their Stringy/Brane Origin", hep-th/9710239;
\next
Cherkis, S.A., A. Kapustin, ``Singular Monopoles and Supersymmetric Gauge
Theories in Three Dimensions", hep-th/9711145.

\item{[24]} Kac, V., {\it ``Infinite-dimensional Lie algebras"},
Birkh\"auser (1983) Boston;
\next
Goddard, P. and D. Olive, ``Kac-Moody and Virasoro algebras
in relation to quantum physics", International J. of Modern Physics A,
Vol. I (1986) 303-414;
\next
McKay, W.G., J. Patera and D.W. Rand, ``Tables of Representations
of Simple Lie Algebras'', Vol. I : Exceptional Simple Lie algebras,
Centre de Math\'ematiques, Universit\'e de Montr\'eal, 1990.

\item{[25]} Erdelyi, A., ed., {\it ``Higher Transcendental Functions''},
Bateman Manuscript Project, Vol. II, R.E. Krieger (1981) Florida.

\item{[26]} Montonen, C., and D. Olive, ``Magnetic Monopoles as Gauge 
Particles ?", Phys. Lett. {\bf B72} (1977) 117.

\item{[27]} 
Vafa, C. and E. Witten, ``Strong Coupling Test of S-Duality",
Nucl. Phys. {\bf B431} (1994) 3, hep-th/9408074;
\next
Minahan, J.A., D. Nemeschansky, ``N=2 Super-Yang-Mills and 
Subgroups of SL(2,Z)", Nucl. Phys. {\bf B468} (1996) 72;
\next
J.A. Minahan, D. Nemeschansky, N.P. Warner, ``Instanton Expansions for 
Mass Deformed N=4 SuperYang-Mills Theories, hep-th/9710146.

\item{[28]} 
Braden, H.W. ``R-Matrices, Generalized Inverses and Calogero Moser Sutherland 
Models", to appear in the {\it Proceedings of the Workshop on
Calogero-Moser-Sutherland Models}, in the CRM Series in Mathematical Physics,
Springer-Verlag; available from http://www.maths.ed.ac.uk/preprints/97-017.

\item{[29]}
Naculich, N.G., H. Rhedin, H.J. Schnitzer, ``One Instanton Predictions of
a Seiberg-Witten Curve from M-Theory : The Anti Symmetric Representation
of SU(N)", hep-th/9804105;
\next
Ennes,I.P., S.G. Naculich, H. Rhedin, H.J. Schnitzer, ``One Instanton Tests of
a Seiberg-Witten Curve from M-Theory : The Symmetric Representation
of SU(N)", hep-th/9804151.

\end